\documentclass[aps,prl,twocolumn,preprintnumbers,superscriptaddress,amsmath,amssymb]{revtex4}

\usepackage{graphicx}
\usepackage{subfigure}
\usepackage{epsfig}
\usepackage{xcolor}
\usepackage{dcolumn}
\usepackage{bm}
\usepackage{ulem}
\usepackage{color}
\usepackage{amsmath}
\usepackage{amsfonts}
\usepackage{amssymb}

\usepackage{lineno}
\usepackage{braket}

\usepackage{hyperref}
\hypersetup{%
	citecolor=[RGB]{0,0,200},%
	colorlinks=true,%
	linkcolor=[RGB]{0,0,200}%
}	

\usepackage{makecell}

\begin{document}

\title{Floquet-Volkov interference in a semiconductor}

\author{Changhua Bao}
\affiliation{Department of Physics, Tsinghua University, Beijing 100084, People's Republic of China}
\affiliation{State Key Laboratory of Low-Dimensional Quantum Physics, Tsinghua University, Beijing 100084, People's Republic of China}

\author{Haoyuan Zhong}
\affiliation{Department of Physics, Tsinghua University, Beijing 100084, People's Republic of China}
\affiliation{State Key Laboratory of Low-Dimensional Quantum Physics, Tsinghua University, Beijing 100084, People's Republic of China}

\author{Benshu Fan}
\affiliation{Department of Physics, Tsinghua University, Beijing 100084, People's Republic of China}
\affiliation{State Key Laboratory of Low-Dimensional Quantum Physics, Tsinghua University, Beijing 100084, People's Republic of China}

\author{Xuanxi Cai}
\affiliation{Department of Physics, Tsinghua University, Beijing 100084, People's Republic of China}
\affiliation{State Key Laboratory of Low-Dimensional Quantum Physics, Tsinghua University, Beijing 100084, People's Republic of China}

\author{Fei Wang}
\affiliation{Department of Physics, Tsinghua University, Beijing 100084, People's Republic of China}
\affiliation{State Key Laboratory of Low-Dimensional Quantum Physics, Tsinghua University, Beijing 100084, People's Republic of China}

\author{Shaohua Zhou}
\affiliation{Department of Physics, Tsinghua University, Beijing 100084, People's Republic of China}
\affiliation{State Key Laboratory of Low-Dimensional Quantum Physics, Tsinghua University, Beijing 100084, People's Republic of China}

\author{Tianyun Lin}
\affiliation{Department of Physics, Tsinghua University, Beijing 100084, People's Republic of China}
\affiliation{State Key Laboratory of Low-Dimensional Quantum Physics, Tsinghua University, Beijing 100084, People's Republic of China}

\author{Hongyun Zhang}
\affiliation{Department of Physics, Tsinghua University, Beijing 100084, People's Republic of China}
\affiliation{State Key Laboratory of Low-Dimensional Quantum Physics, Tsinghua University, Beijing 100084, People's Republic of China}

\author{Pu Yu}
\affiliation{Department of Physics, Tsinghua University, Beijing 100084, People's Republic of China}
\affiliation{State Key Laboratory of Low-Dimensional Quantum Physics, Tsinghua University, Beijing 100084, People's Republic of China}

\author{Peizhe Tang}
\email{peizhet@buaa.edu.cn}
\affiliation{School of Materials Science and Engineering, Beihang University, Beijing 100191, People's Republic of China}
\affiliation{Max Planck Institute for the Structure and Dynamics of Matter, Center for Free Electron Laser Science, 22761 Hamburg, Germany}

\author{Wenhui Duan}
\affiliation{Department of Physics, Tsinghua University, Beijing 100084, People's Republic of China}
\affiliation{State Key Laboratory of Low-Dimensional Quantum Physics, Tsinghua University, Beijing 100084, People's Republic of China}
\affiliation{Frontier Science Center for Quantum Information, Beijing 100084, People's Republic of China}
\affiliation{Institute for Advanced Study, Tsinghua University, Beijing 100084, People’s Republic of China}

\author{Shuyun Zhou}
\email{syzhou@mail.tsinghua.edu.cn}
\affiliation{Department of Physics, Tsinghua University, Beijing 100084, People's Republic of China}
\affiliation{State Key Laboratory of Low-Dimensional Quantum Physics, Tsinghua University, Beijing 100084, People's Republic of China}
\affiliation{Frontier Science Center for Quantum Information, Beijing 100084, People's Republic of China}

\date{\today}

\begin{abstract}

Intense light-field can dress both Bloch electrons inside crystals and photo-emitted free electrons in the vacuum, dubbed as Floquet and Volkov states respectively. These quantum states can further interfere coherently, modulating light-field dressed states. Here, we report experimental evidence of the Floquet-Volkov interference in a semiconductor - black phosphorus. A highly asymmetric modulation of the spectral weight is observed for the Floquet-Volkov states, and such asymmetry can be further controlled by rotating the pump polarization. Our work reveals the quantum interference between different light-field dressed electronic states, providing insights for material engineering on the ultrafast timescale.
\end{abstract}

\maketitle

\begin{figure}[t]
	\centering
	\includegraphics[]{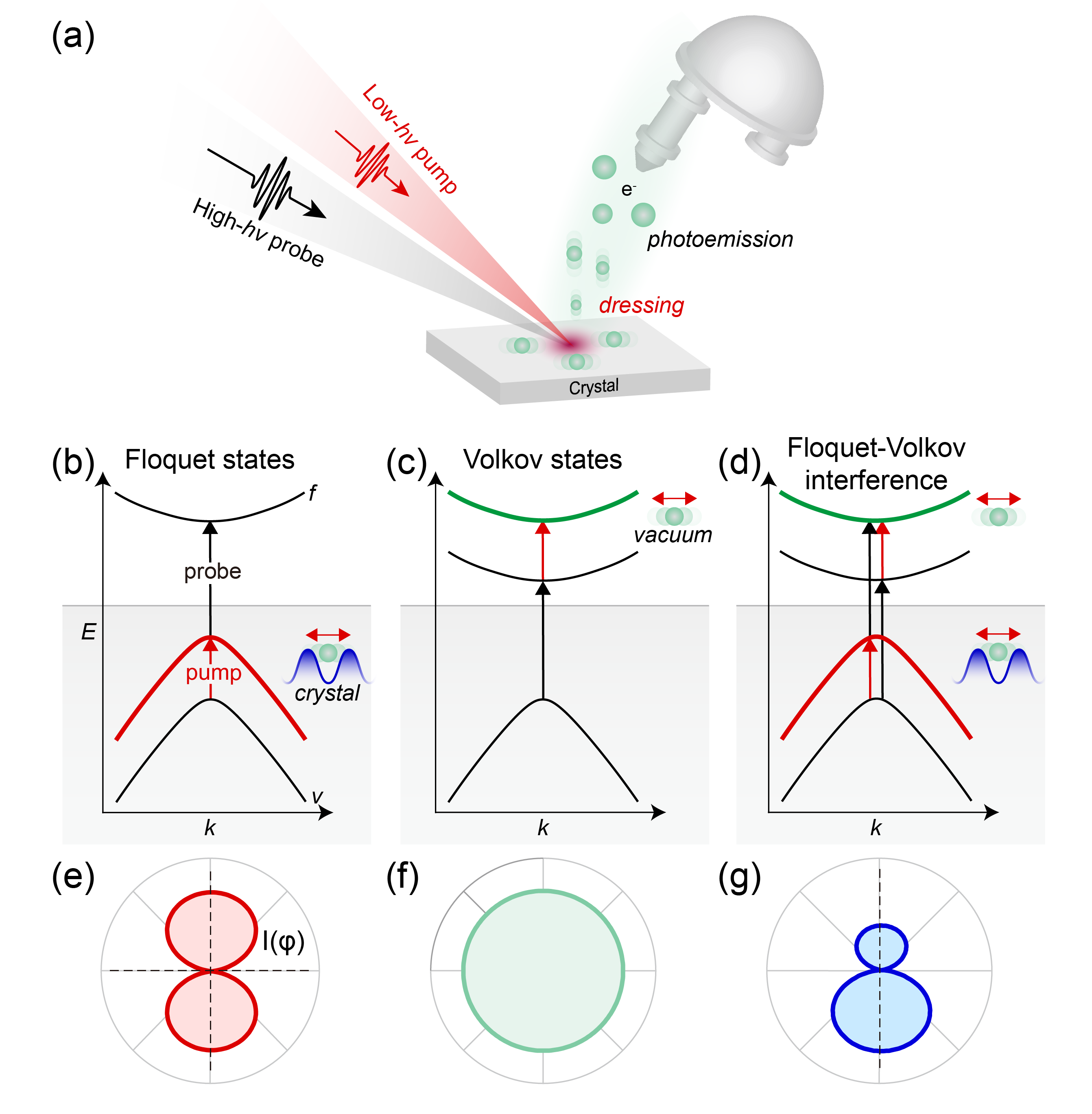}
	\caption{(a) The schematic for TrARPES with pump and probe lights to reveal Floquet-Volkov interference. The green balls with shadows represent the light-dressed electrons in crystal and vacuum. (b-d) The schematics for illustrating pump and probe lights induced optical transitions for Floquet, Volkov, and Floquet-Volkov states. The valence band is denoted by $v$, and the free-electron final state is denoted by $f$. (e-g) The angular distribution of the relative spectral weight of the first-order light-dressed sideband ($n$ = 1) with respect to the original band ($n$ = 0) for Floquet, Volkov, and Floquet-Volkov states.} 
	\label{fig:Intro} 
\end{figure}

\begin{figure*}[t]
	\centering
	\includegraphics[]{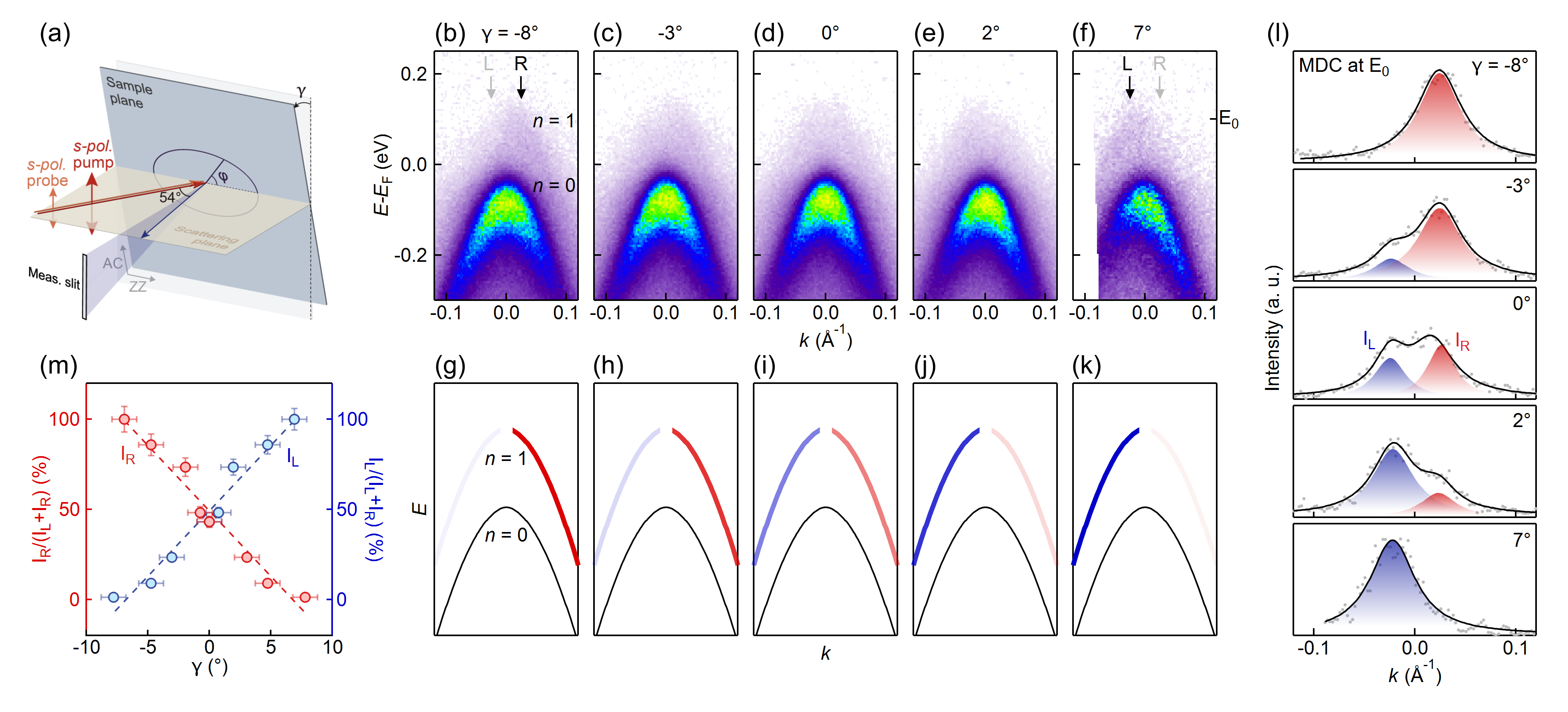}
	\caption{(a) The schematic for experimental geometry and sample rotation. (b-f) TrARPES dispersion images along the armchair (AC) direction measured at different $\gamma$ angles. (g-k) Schematics of the asymmetric sidebands with different $\gamma$. (l) Extract momentum distribution curves from the data in (b-f) at the energy of $E_0$ as indicated in (f). (m) Normalized spectral weight of the left and right branches as a function of $\gamma$.  The amplitudes of the two branches are obtained from the fitting in (l), denoted by $I_L$ and $I_R$, respectively. The normalized spectral weights of the two branches are defined by $I_R/(I_L+I_R)$ and $I_L/(I_L+I_R)$.  The pump photon energy is 160 meV, and the pump fluence is 0.6 mJ/cm$^{2}$.} 
	\label{fig:tilting} 
\end{figure*}

\begin{figure}[t]
	\centering
	\includegraphics[scale=1.0]{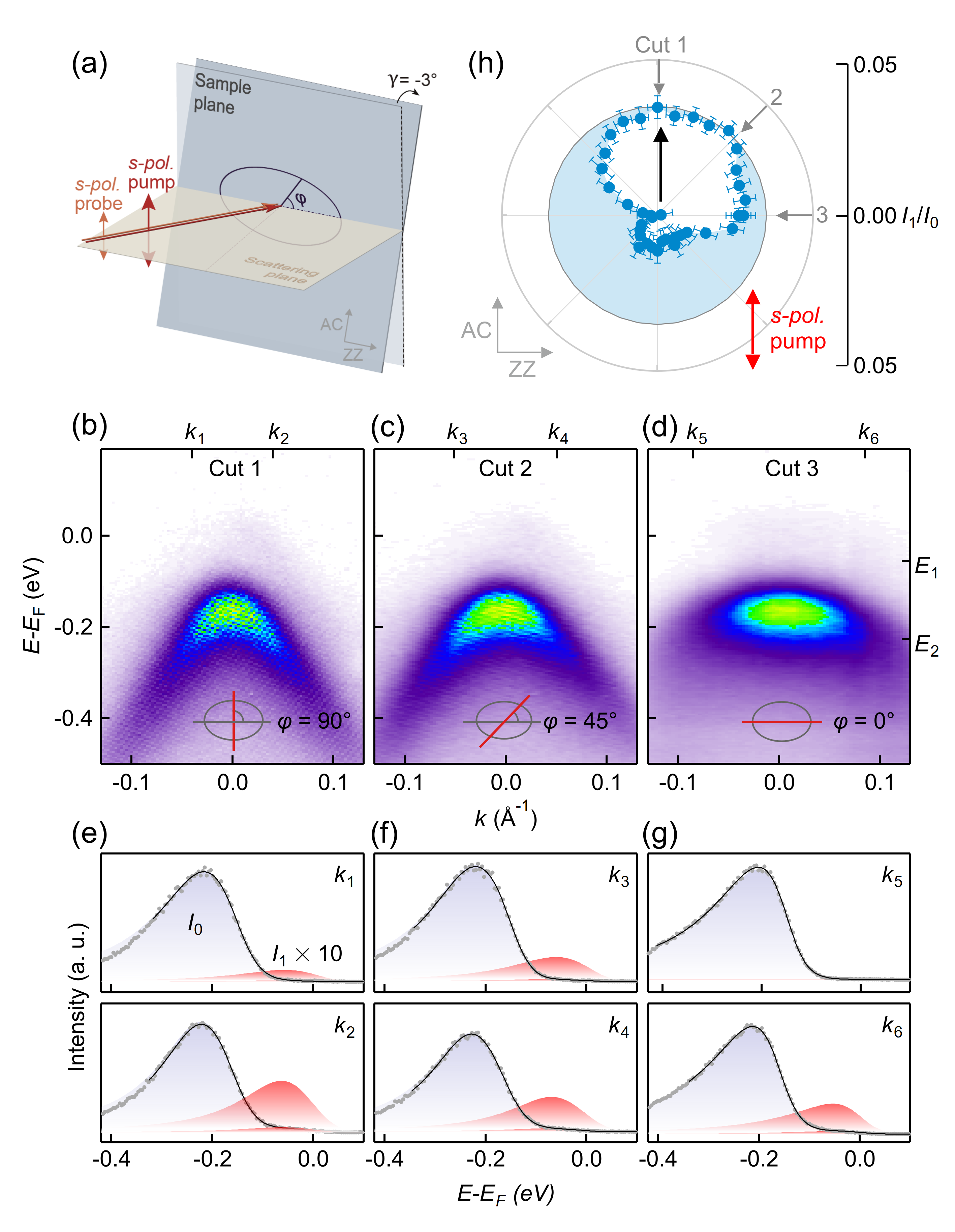}
	\caption{ (a) The schematic for experimental geometry with $\gamma = -3^{\circ}$. (b-d) TrARPES dispersion images cutting through different directions as indicated in the inset. (e-g) Energy distribution curves at different momenta as indicated in (b-d), which correspond to the same energy position $E_1$ and $E_2$. (h)  The polar plot of extracted normalized spectral weight of the $n$ = 1 sideband. The pump photon energy is 160 meV, and the pump fluence is 0.6 mJ/cm$^{2}$.}
	\label{fig:sMap} 
\end{figure} 

\begin{figure}[t]
	\centering
	\includegraphics[scale=1.0]{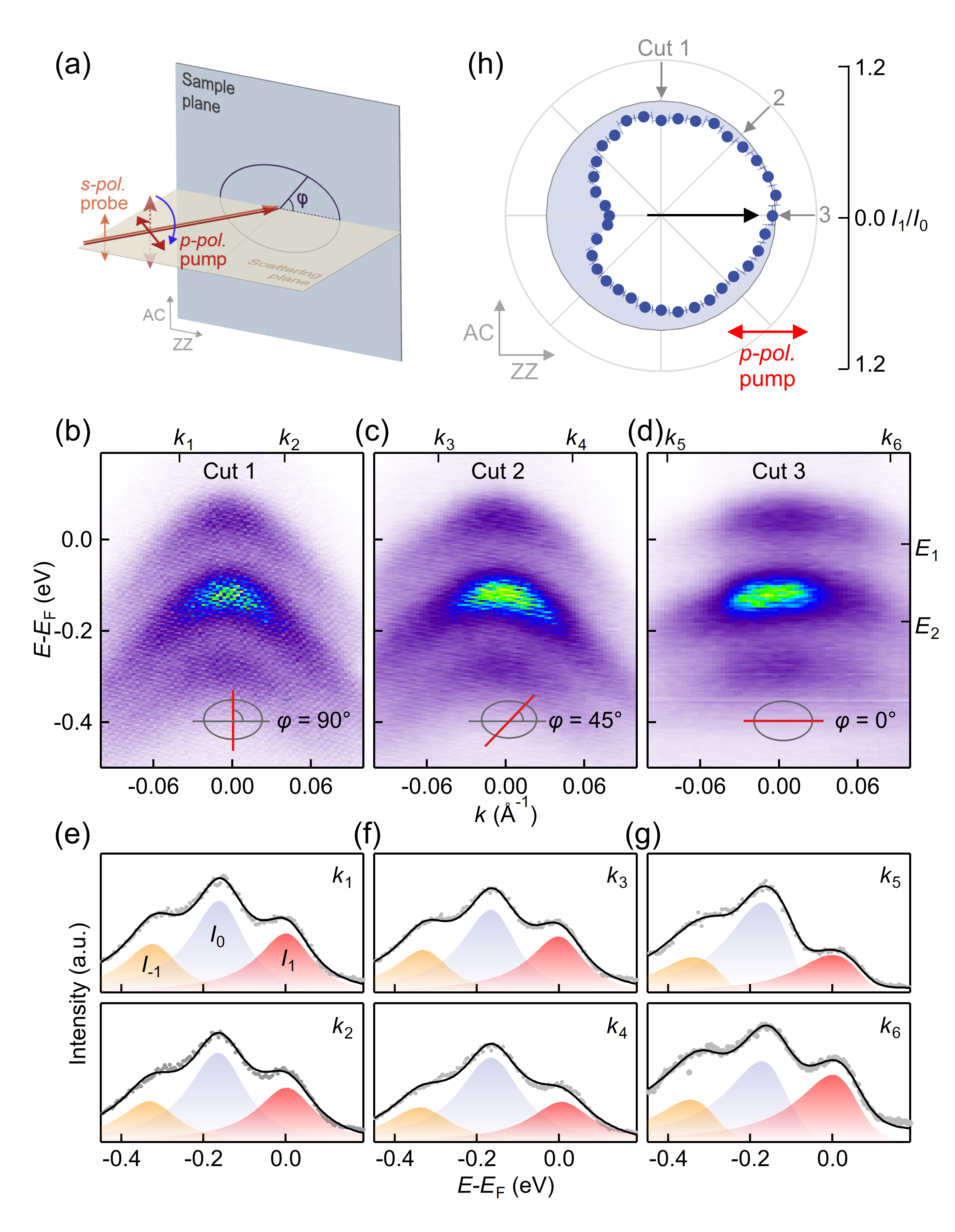}
	\caption{ (a) The schematic for experimental geometry with $p$-$pol.$ pump and an incident angle $\theta = 54^{\circ}$. (b-d) TrARPES dispersion images cutting through different directions as indicated in the insets. (e-g) EDCs at different momenta as indicated in (b-d), which corresponds to the same energy position $E_1$ and $E_2$. (h) The polar plot of extracted normalized spectral weight of the $n$ = 1 sideband. The pump photon energy is 160 meV, and the pump fluence is 0.4 mJ/cm$^{2}$.} 
\label{fig:pMap} 
\end{figure}

\begin{figure}[t]
	\centering
	\includegraphics[width = 9cm]{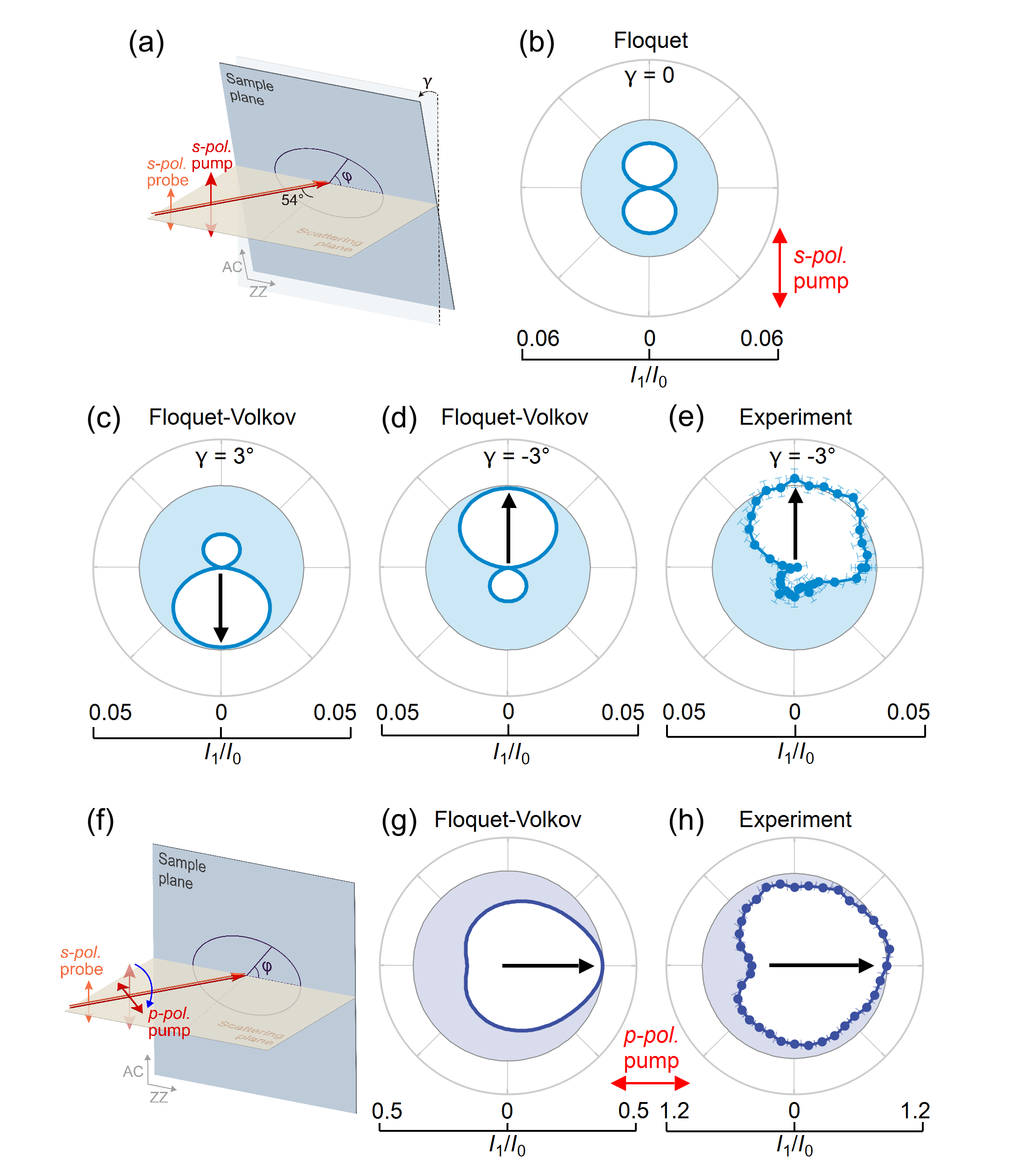}
	\caption{ (a)  The schematic for experimental geometry for $s$-$pol.$ pump. (b-d) Polar plot of the simulated angular distribution of the normalized spectral weight of $n$ = 1 sideband ($I_1/I_0$) at different $\gamma$ angles (0$^{\circ}$ and $\pm3^{\circ}$) based on Floquet-Volkov interference. (e) Polar plot of the experimental normalized spectral weight of $n$ = 1 sideband upon $s$-$pol.$ pump at $\gamma = -3^{\circ}$. (f) The schematic for experimental geometry for $p$-$pol.$ pump. (g) Polar plot of the simulated angular distribution of the normalized spectral weight of $n$ = 1 sideband for $p$-$pol.$ pump. (h) Polar plot of the experimental normalized spectral weight of $n$ = 1 sideband upon $p$-$pol.$ pump. } 
\label{fig:cal}
\end{figure}

Light can interact with electrons in quantum materials via different coupling mechanisms \cite{basov2017towards,okaRev2019,Lindner2020NRP,Chris2021rev,Sentef2021RMP,ZhouNRP2021,bloch2022strongly}. For example, photo-electrons can be emitted from solid-state materials into the vacuum when the photon energy is larger than the work function, and such photoelectric effect \cite{hertz1887ueber,einstein1905heuristic} is one of the most important foundations for Quantum Mechanics. When the light-field is strong enough which is more easily achievable for low photon energy, the strong light-matter interaction can lead to light-field dressed electronic states, for example,  light-field dressed Bloch electrons in crystals dubbed as Floquet states [see Fig.~\ref{fig:Intro}(b)] \cite{Oka2009PhotoHE,Demler2011PhotoHE,Lindner2011natphy,Gedik2013Sci,Gedik2015NM,Gedik2018Science,Cavalleri2020NP,Hsieh2021Nature,aeschlimann2021survival,Lee2022Nature,Zhou2023Nature,ZhouPRL2023,Huber2023Nature,Shambhu2023NP,Kogar2024NM,Zhou2024ACS,Zhou2024NC}, and light-dressed free electrons in the vacuum dubbed as Volkov states [see Fig.~\ref{fig:Intro}(c)] \cite{wolkow1935klasse,madsen2005strong}. Both of these two quantum states can be detected in time- and angle-resolved photoemission spectroscopy (TrARPES) experiments [see Fig.~\ref{fig:Intro}(a)]. For example, the Bloch electrons in the crystal can be dressed by a low-energy pump pulse to form Floquet states, which are then excited to the vacuum by a high-energy probe pulse. On the other hand, the photo-excited free electrons by the probe pulse in the vacuum can also be dressed by the low-energy pump pulse to form Volkov states as schematically illustrated in Fig.~\ref{fig:Intro}(d). These different quantum states can further interfere, i.e., Floquet-Volkov interference \cite{Park2014PRA}, leading to modulation of light-field induced sidebands.

So far, the Floquet-Volkov interference has been reported in the two-dimensional (2D) Dirac cone surface state of topological insulator Bi$_2$Se$_3$ \cite{Gedik2016NP}, where the spectral weight of photon-dressed sideband shows distinct angular distribution upon pumping with different light polarizations. 
More recently, Floquet-Volkov interference has been observed in another Dirac material - graphene \cite{Stefan2024Graphene,Nuh2024Graphene}, where the effect of the interference on the spectral weight modulation is used to support the formation of the Floquet states. How such Floquet-Volkov interference modulates the spectral weight of light-field dressed states in non-Dirac materials is an interesting and important question to answer, because it can provide insights for extending Floquet states to a wider range of materials. 

On the other hand, photoemission spectroscopy, as the standard technique for the electronic structure and spectral functions, is well understood in terms of quasiparticles, their lifetimes and interactions in the past decades \cite{damascelli2003angle,ZXRMP21}. However, the involvement of both bound and free electrons makes TrARPES a much more complex dynamic process and cannot correspond to the intrinsic dynamics of quantum materials directly.  Moreover, a deeper understanding is also needed for coherent interaction between light and free electrons \cite{kfir2020controlling,wang2020coherent,reutzel2020coherent} the laser-driven electron sources \cite{wimmer2014terahertz,green2018bright,li2023coherent}.

Herein, by performing TrARPES measurements on a semiconducting black phosphorus with low pump photon energy which is favorable for inducing Floquet/Volkov states, we reveal the effects of Floquet-Volkov interference on the light-dressed electronic states. The Floquet-Volkov interference is evident by the asymmetry of the spectra weight distribution of the sideband, and more importantly, we find that such asymmetry can be further manipulated by tuning the polarization of the driving light-field.

Black phosphorus is a layered material and a narrow-gap semiconductor. The high mobility and the simple band structure featuring a parabolic-like conduction band (CB) and valence band (VB) at the $\Gamma$ point \cite{JiNC2014,XiaNC2014,LiBP2014,JiaBPARPESPRB2014} make it an ideal material for investigating Floquet engineering. Recently, light-induced Floquet band engineering has been reported in black phosphorus upon near-resonance \cite{Zhou2023Nature,Zhou2024ACS} and below-gap pumping \cite{ZhouPRL2023,Zhou2024NC}. Taking the coupling between linear polarized light and the electronic states of black phosphorus on the band edge [see Supplemental Material (SM) for more details \cite{supp}] into account, a simple calculation shows that the spectral weight of the Floquet states with Floquet index $n$ = 1 exhibits gourd shape pattern in the momentum space which is symmetric about the vertical and horizontal axes [see Fig.~\ref{fig:Intro}(e)], while the Volkov sideband exhibits an isotropic pattern [see Fig.~\ref{fig:Intro}(f)]. Interestingly, although the pure Floquet and Volkov states maintain a symmetric distribution of the spectral weight for the sideband, their interference leads to a highly asymmetric distribution of the spectral weight about the horizontal axis [see Fig.~\ref{fig:Intro}(g)].  It is worth noting that the above discussion is valid around the $\Gamma$ point where the contribution of in-plane light-field can be neglected for Volkov states \cite{keunecke2020electromagnetic}. In this work, we focus on the experimental evidence of such Floquet-Volkov interference and how the light-field can be further used to manipulate such anisotropic modulation.

TrARPES measurements were performed with low pump photon energy down to 160 meV to enhance the Floquet and Volkov states \cite{ZhouPRL2023}. The experimental geometry is shown in Fig.~\ref{fig:tilting}(a): the pump and probe pulses are both $s$-polarized ($s$-$pol.$) and the polarization is along the armchair (AC) direction of black phosphorus at an incident angle of $54^{\circ}$, and the measurement direction is also along the AC direction. The Floquet-Volkov interference can be revealed by the asymmetry when rotating the pump polarizations, or rotating the sample angle while keeping the same pump polarization.

To reveal the Floquet-Volkov interference, the sample is tilted by a $\gamma$ angle as schematically illustrated in Fig.~\ref{fig:tilting}(a). Such tilt angle effectively changes the pump light-field and introduces an out-of-plane light-field, which is needed for introducing the Volkov states \cite{Gedik2016NP}. Such change in the effective pump field does not change the dispersion of the light-induced sideband; however, an interesting asymmetric spectral weight distribution of the Floquet sideband $n$ = 1 is observed in Fig.~\ref{fig:tilting}(b), where the right branch is stronger than the left branch, as pointed by the arrows. Such asymmetry indicates that the mirror symmetry of the whole system about the scattering plane is broken by the tilting. Moreover, the asymmetry is found to be tunable by rotating the $\gamma$ angle. By rotating the $\gamma$ angle from negative value to positive value in Fig.~\ref{fig:tilting}(b-f), the asymmetry becomes opposite: the left branch is stronger than the right branch [Fig.~\ref{fig:tilting}(f)]. These results demonstrate a tunable asymmetric spectral weight distribution of the light-induced sideband as schematically summarized in Fig.~\ref{fig:tilting}(g-k). The spectral weight of right and left branches extracted from the momentum distribution curves (MDCs) at E$_0$  [Fig.~\ref{fig:tilting}(l)], where two well-separated peaks without disturbing from the original band (n = 0) are distinguished, allowing a straightforward analysis of the two branches. The extracted spectral weight further shows that an asymmetry as high as 100\% can be achieved just by tilting a small $\gamma$ angle less than 10$^{\circ}$ at this energy [Fig.~\ref{fig:tilting}(m),  see Fig.~S1,2 for more analysis at different energies and corresponding calculation in SM \cite{supp}], indicating that the intensity of the sideband is strongly tunable by the pump field. Such intensity asymmetry after introducing the Volkov states (by changing the tilt angle)  is a signature of the Floquet-Volkov interference.

The above results demonstrate the Floquet-Volkov interference along a one-dimensional (1D) momentum line-cut. To reveal the effect of the Floquet-Volkov interference in full two-dimensional momentum space, the 2D TrARPES spectrum is measured with a nonzero $\gamma$ angle [$\gamma$ = -3$^\circ$, see Fig.~\ref{fig:sMap}(a)]. Figure~3(b-d) shows three representative TrARPES dispersion images extracted along different directions as indicated in the insets [$\varphi$ = 90$^{\circ}$, 45$^{\circ}$ and 0$^{\circ}$, where $\varphi$ is the in-plane distribution angle defined in Fig.~\ref{fig:sMap}(a)]. The spectral weight of $n$ = 0 VB ($I_0$) and $n$ = 1 sideband ($I_1$) is extracted from the energy distribution curves (EDCs) in Fig.~\ref{fig:sMap}(e-g) (see Fig.~S3 for the full data in SM \cite{supp}). The normalized spectral weight ($I_1/I_0$) as a function of $\varphi$ is plotted as a polar plot in Fig.~\ref{fig:sMap}(h), which shows that the main spectral weight anisotropy is along the AC direction as indicated by the black arrow, consistent with the observed asymmetry in the measurement along the AC direction in Fig.~\ref{fig:tilting}. The asymmetry in the normalized spectral weight ($I_1/I_0$) is dominantly contributed by the change in the $n$ = 1 sideband while the change in the intensity of the valence band ($I_0$) is much less (see details in Fig.~S4 of SM \cite{supp}), suggesting that it is mostly caused by the Floquet-Volkov interference. Interestingly, the anisotropic direction marked by the black arrow in Fig.~\ref{fig:sMap}(h) is along the pump polarization direction marked by the red arrow in Fig.~\ref{fig:sMap}(h), implying that the anisotropy might depend on the pump light-field direction.

To further explore how the spectral weight anisotropy is related to the pump light-field direction, the pump polarization is rotated by $90^{\circ}$ to $p$-$pol.$ pump [see Fig.~\ref{fig:pMap}(a)]. Due to the presence of an out-of-plane light-field for the $p$-$pol.$ pump, strong sidebands are observed in Fig.~\ref{fig:pMap}(b-d) due to contributions from the Volkov states. The corresponding spectral weight is extracted from the EDCs in Fig.~\ref{fig:pMap}(e-g) (see Fig.~S5 for the full data in SM \cite{supp}). The spectral weight distribution of $n$ = 1 sideband is strikingly different from the $s$-$pol.$ pump case. The extracted $I_1/I_0$ in Fig.~\ref{fig:pMap}(h) is symmetric about the horizontal axis but asymmetric about the vertical axis as indicated by the black arrow, which is also dominated by the $n$ = 1 sideband (see details in Fig.~S6 of SM \cite{supp}). The anisotropic direction is rotated by $90^{\circ}$ marked by the black arrow in Fig.~\ref{fig:pMap}(h), which is the same as the pump light-field direction marked by the red arrow in Fig.~\ref{fig:pMap}(h), thus suggesting that spectral weight anisotropy of the light-induced sidebands strongly depends on the pump light-field direction.

To further support that the observed spectral weight anisotropy originates from the Floquet-Volkov interference, we apply the Floquet theory to calculate the wavefunctions for the Floquet states in black phosphorus and the Volkov states for free electrons in the vacuum, and to simulate the Floquet-Volkov interference correspondingly (see SM for more details \cite{supp}).  Under the Peierls substitution \cite{PhysRevB.14.2239}, the wavefunctions of the Volkov and Floquet states can be written as functions of the light-matter interaction parameters $\alpha$ and $\beta$ respectively, herein $\alpha$ depends on the out-of-plane light-field of the pumping laser and $\beta$ can be modulated by in-plane polarization direction of pumping light-field. When there is Floquet-Volkov interference, the photoemission intensity is determined by both $\alpha$ and $\beta$. For the $s$-$pol.$ pump case in Fig.~\ref{fig:cal}(a), different $\gamma$ angle results in different in-plane and out-of-plane components of pump light-fields, and the different values of $\alpha$ and $\beta$ results in an asymmetric intensity pattern of photoemission for the light-induced sidebands. This is the physical origin of the observed distinct spectral weight distribution. For $\gamma = 0$, the angular distribution of the normalized spectral weight of $n$ = 1 sideband ($I_1/I_0$) shows a symmetric gourd shape pattern in Fig.~\ref{fig:cal}(b), following the direction of the pump light-field for pure Floquet states. When the Floquet-Volkov interference is included by introducing an out-of-plane light-field component via rotating to a non-zero $\gamma$, the pattern becomes asymmetric along the vertical direction as shown in Fig.~\ref{fig:cal}(c,d), which is qualitatively consistent with the experimental asymmetric feature shown in Fig.~\ref{fig:cal}(e). When reversing the sign of the $\gamma$ angle, the relative phase between out-of-plane and in-plane light-fields changes, resulting in the sign change of $\alpha$ in the Floquet-Volkov coefficient (see Eq.~7 in SM \cite{supp}) and the asymmetric pattern of the normalized spectral weight [see Fig.~\ref{fig:cal}(d,e)]. In short, the asymmetric direction is determined by the relative phase between Floquet and Volkov states, which is dictated by the relative phase between out-of-plane and in-plane light-fields.

Moreover, by rotating the pump polarization from $s$-$pol.$ pump to $p$-$pol.$ pump [Fig.~\ref{fig:cal}(f)], the asymmetric direction rotates with the pump polarization as indicated by the black arrow in Fig.~\ref{fig:cal}(g), herein our simulations well reproduce the experimental asymmetric feature qualitatively as shown in Fig.~\ref{fig:cal}(h). The simulation shows that the direction and magnitude of spectral weight anisotropy of the first-order sideband can be controlled by the pump polarization and pointing direction of the light-field, respectively.

In summary, we report the Floquet-Volkov interference in semiconducting black phosphorus with nearly parabolic electronic structures, where such interference is evident by a novel anisotropic distribution of the spectral weight of the first-order ($n$ = 1) light-induced sideband. Moreover, an efficient manipulation of such spectral weight anisotropy can be achieved by controlling the polarization of the pumping light-field with respect to the crystal. Such effects have been used to distinguish the relatively weak Floquet states from the strong Volkov states for the surface states of a topological insulator \cite{Gedik2016NP} as well as graphene  \cite{Stefan2024Graphene,Nuh2024Graphene}. In the case of black phosphorus, while the contribution of Floquet states has already been established by the Floquet band engineering \cite{Zhou2023Nature,ZhouPRL2023,Zhou2024ACS,Zhou2024NC}, resolving and understanding such Floquet-Volkov interference is still quite useful, as it can provide insights for extending Floquet states into other non-Dirac materials. In particular, considering that Floquet states have been reported only in a few materials, the understanding of Floquet-Volkov interference in black phosphorus could be used to identify Floquet states in other non-Dirac materials where light-induced band engineering (modification of the electronic structure) may not be strong enough to be detected. Moreover, our work not only provides a deeper understanding of the process of TrARPES, where the electrons are tailored by both probe light and the pump light geometry, but also a new degree of freedom to manipulate and design the laser-driven free-electron sources on ultrafast timescales.

\begin{acknowledgments}

This work is supported by the National Natural Science Foundation of China (Grant Nos.~12234011, 52388201, 12421004, 92250305, 12374053), National Key R\&D Program of China (Grant Nos.~2021YFA1400100, 2020YFA0308800, 2024YFA1409100) and New Cornerstone Science Foundation through the XPLORER PRIZE. Changhua Bao acknowledges support from the Project funded by China Science Foundation (Grant No. BX20230187) and the Shuimu Tsinghua Scholar Program.

\end{acknowledgments}


\end{document}